# Fast Security Evaluation for Operation of Water Distribution Systems Against Extreme Conditions *

Mostafa Goodarzi, Dan Wu, *Member, IEEE*, and Qifeng Li, *Senior Member, IEEE*

*Abstract—* This paper defines a security injection region (SIR) to guarantee reliable operation of water distribution systems (WDS) under extreme conditions. The model of WDSs is highly nonlinear and nonconvex. Understanding the accurate SIRs of WDSs involves the analysis of nonlinear constraints, which is computationally expensive. To reduce the computational burden, this paper first investigates the convexity of the SIR of WDSs under certain conditions. Then, an algorithm based on a monotone inner polytope sequence is proposed to effectively and accurately determine these SIRs. The proposed algorithm estimates a sequence of inner polytopes that converge to the whole convex region. Each polytope adds a new area to the SIR. The algorithm is validated on two different WDSs, and the conclusion is drawn. The computational study shows this method is applicable and fast for both systems.

## I. Introduction

Water security is generally defined by water consumers' accessibility to an adequate quantity of water with acceptable quality. One of the fundamental issues in water security is the imbalance between supply and demand [1]-[2], which is subject to the water pipe limits. The ability of a water distribution system (WDS) to supply the consumers with water in the minimally acceptable quantity and quality, under normal and abnormal conditions, is termed as network reliability [4]-[7]. The reliability of WDS can be studied in two different groups: mechanical failures from outages due to natural and man-made disasters, extreme weather events, and aging water infrastructures, and hydraulic failures such as the inappropriate operation of pumps due to network tolerance for the demand change, and cost estimation [4]-[8].

To reduce the energy cost of WDS operations, the Optimal Pump Scheduling (OPS) problems have been investigated for decades [9]-[16]. In [13], the mixed-integer nonlinear programming problem of WDS was relaxed and converted to a mixed-integer convex program by developing a quasi-convex hull relaxation. In [14], the OPS problem was solved by a hybrid optimization model that uses linear programming approximation and a greedy algorithm. The authors of [15] proposed a mixed-integer second-order cone relaxation for WDS to solve the OPS problem and focus on energy cost reduction in the WDS. In [16], an ant colony optimization framework was used to solve the OPS problems. All of OPS models have assumed water demand at nodes are constant value in each time slot and obtained one feasible solution.

The OPS problem focuses on finding the optimal operation schedules of pumps assuming that the nodal demand profiles are given via prediction. However, in real-world applications, a water consumer may need more or less water than the forecasted value. Indeed, the water demands change from time to time and sometimes are quite different from the ahead-of-real-time predictions. For example, an extreme case such as a forest fire, will change the nodal water demand completely and affect the feasibility of an OPS solution which is based on forecasted water demands. Therefore, evaluating the security of WDSs is an important issue for the demand side, especially, in an extreme situation.

It is believed that water and power systems are tightly intertwined [17]-[21]. The pumps and most of the water facilities require electricity to work. For instance, water networks consume 4% to 16% of the total electricity consumption in the United States [17]. On the other hand, water usage is necessary for generating electric power and cooling power plant. The WDS and the power systems are large complex networks, and combining them results in a more challenging problem. Several studies investigate the co-optimization of these systems [18]-[20]. However, they do not consider WDS operations to meet nodal water demand under emergency conditions. If WDS could not meet the power plant cooling water demand, the power system could not generate enough power for pumping stations in WDS, and a cascading failure will happen for both systems [21]. Besides, in real applications, interconnected systems' conflicts prevent their actual real-time models, such as pipe network information, from being shared in a co-optimization. As a result, evaluations of the Security Injection Region (SIR) against extreme conditions with simple descriptions are more practical for sharing the information among such systems.

This paper defines an SIR to capture the whole range of the secure operation of WDS. A WDS operator can provide the SIR as a guideline for all customers to determine the whole feasible region for water demand, and customers can evaluate the security of WDS to know the acceptable value for their water demands. A water demand value located outside the SIR can cause a hydraulic pressure failure in the WDS operation. Real-time water demand is related to the demand side and, WDS cannot control it. Hence real-time

*Goodarzi and Li's work was supported by U.S. National Science Foundation under grant#1808988, while Wu's work was supported by the MI-MIT Cooperative Program under grant MM2017-000002.
M. Goodarzi and Q. Li are with the Department of Electrical and Computer Engineering, University of Central Florida, Orlando, Fl 32816 USA (e-mail: mostafa.goodarzi@knights.ucf.edu, Qifeng.li@ucf.edu).
Dan Wu is with the Lab for Information and Decision Systems at Massachusetts Institute of Technology, Cambridge, MA 02139. (e-mail: danwumit@mit.edu).

water demand should be located in the SIR, by customers, to avoid any hydraulic failures.

However, obtaining an accurate SIR of the WDS is computationally challenging since it involves the analysis of highly nonlinear constraints. This paper aims at effectively determining and finding an SIR. For this purpose, we leverage a method of monotone inner polytope sequence, which was proposed in our prior work for gas networks [22], to calculate the SIRs of water networks.

To the best of our knowledge, this is the first time a study addresses the SIR of WDS where all feasible solutions of unpredictable water demands in abnormal situations are determined at once. A method as sketched in Fig. 1, is proposed to evaluate the SIR and guarantee the security of WDS. There are two main parts in our proposed method: the WDS part (blue rectangle) and the customer part (red rectangle). Initially, an optimization algorithm is applied to approximate the convex region for uncertain nodal water demands after running the OPS for normal situations. Each step of the optimization algorithm produces a new polytope in the sequence that is an internal estimation of the region's loadability and guarantees any load profile's feasibility within the polytope. Then, the WDS operator provides the SIR to the demand side, and all customers, especially large costumers like power plants, should control their demand within the SIR. If any customer cannot control its demand properly, the WDS must rerun the OPS with new water demands. The main contributions of this paper are categorized as follows:

- A concept of security injection region is introduced for guaranteeing the security and reliability of WDS operation.
- We show that the SIR is convex under some mild conditions even though the mathematical model of water networks is nonlinear.
- A sequential optimization approach based on the monotone inner polytope sequence is applied to effectively calculating the SIR of WDS.

The rest of this paper is organized as follows. The problem formulation is explained in section II. Section III discusses the solution method as well as the assumptions for convexity of the injection region for steady-state water flow. The algorithm of constructing the monotone inner polytope sequence is introduced in the same section. Two case studies are executed in Section V, while the conclusion is drawn in Section VI.

## II. PROBLEM FORMULATION

We introduce the mathematical model of WDSs for solving the OPS and SIR problems in this section. The constraints such as the pipe network model, pump model, and mass flow conservation law are explained in detail.

### A. Model of Water Distribution System

Pipe Network Model:

The pipe network of a WDS is a directed graph that consists of $\mathcal{N}$ nodes and $\varepsilon$ pipes. The head lost along the pipe can be calculated using different formulas, such as Darcy-Weisbach formula that is the most theoretically accurate [23]. We compute head loss with Darcy-Weisbach equation [24]:

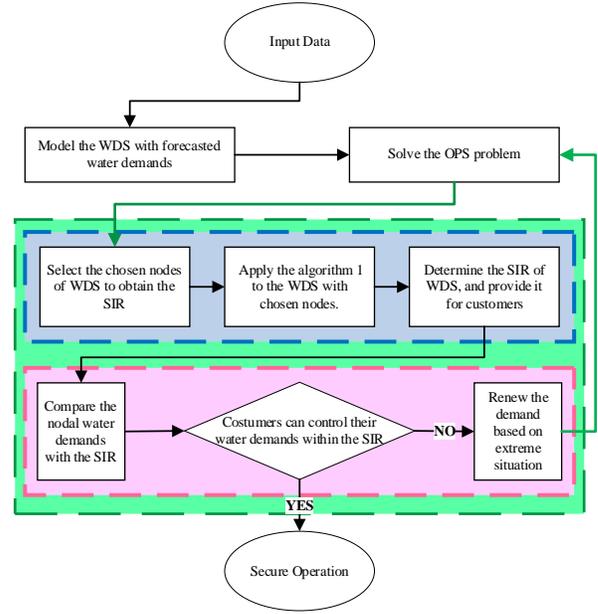

Figure 1. the method to obtain the SIR

$$y_i - y_j + h_i - h_j = R_{ij} sgn(f_{ij}) f_{ij}^2, \quad (1)$$

where $y_i$ and $h_i$ are the water head and elevation of node $i$, and $R_{ij}$ and $f_{ij}$ are head loss coefficient and water flow of the pipe between node $i$ and $j$, respectively. Let $\boldsymbol{Y} = [y_1, \ldots, y_n]^T$ be the column vector of water head at nodes, $A$ be an $|\varepsilon| \times |\mathcal{N}|$ incidence matrix such that $A_{nm} \in \{-1, 0, 1\}$, $\boldsymbol{H} = [h_1, \ldots, h_n]^T$ be the column vector of elevation at nodes, $\boldsymbol{f} = [f_1, \ldots, f_m]^T$ be the column vector of water flow in pipes, $Sgn(f)$ be the $m \times m$ diagonal matrix related to water flow direction, and $M(R)$ be the $m \times m$ diagonal matrix of head loss coefficient of pipes. The pipe network model can be expressed as:

$$AY + AH = M(R)(f \circ f) Sgn(f). \quad (2)$$

Pump Model:

In this paper, we consider that a pump is a special pipe with a head gain imposed. We further assume that pumps are of constant-speed which can be modeled by a quadratic function of the water flow across the pump, i.e., $\frac{y^G}{s^2} = a_2 \left(\frac{f}{s}\right)^2 + a_1 \frac{f}{s} + a_0$ [13], [15], and [25], where $y^G$ and $s$ are head increase across the pump and normalized pump speed, respectively. In this paper, we consider $s$ is equal to 1. $a_2$, $a_1$, and $a_0$ are individual pump head flow coefficients. Since $a_2 f^2 \ll a_1 f + a_0$, the pump model can be reduced to $y^G = a_1 f + a_0$ [13]. As a result, the model of pumps can be formulated as:

$$AY + AH + G = M(R)(f \circ f) Sgn(f), \quad (3)$$

where $\boldsymbol{G} = [y^{G_1}, \ldots, y^{G_m}]^T$ is the column vector of head gains imposed by the pumps. It is assumed that there is no water flowing therein if a pump is off. Therefore, the ON/OFF status of a pump should be considered in the model. Let $S \in \mathbb{R}^{m \times m}$ be a diagonal matrix related to the ON/OFF status of pumps. The diagonal entries of $S$ are 1 for pipes without pump and pipes with the on-pump and, otherwise, zero. Such that, equation (3) will be modified into

$$SAY + SAH + SG = M(R)(f \circ f) Sgn(f). \quad (4)$$

Mass Flow Conservation Law:

The WDS must preserve the total mass flow rate at every node. Several types of pipe would be connected to each node: transfer pipe, inject pipe from the water source, and customer-owned pipe. Let $\boldsymbol{F^G} = [f_1, \ldots, f_n]^T$ be the column vector of water flow inject from the water sources, and $\boldsymbol{d} = [d_1, \ldots, d_n]^T$ be the column vector of water flow to the customer or water demand. The mass flow conservation law can be expressed as:

$$A^T f = F^G - d. \quad (5)$$

Upper and Lower Constraints:

There are several operational and engineering constraints for a WDS. The constraints are defined as follow:

$$y_{min}^G \leq y^G \leq y_{max}^G, \quad (6)$$

$$y_{imin} \leq y_i \leq y_{imax}, \quad (7)$$

$$f_{imin}^G \leq f_i^G \leq f_{imax}^G, \quad (8)$$

$$d_{imin} \leq d_i \leq d_{imax}, \quad (9)$$

$$f_{kmin} \leq f_k \leq f_{kmax}, \quad (10)$$

where $y_{min}$ and $y_{max}$, $d_{imin}$ and $d_{imax}$, and $f_{imin}^G$ and $f_{imax}^G$ are the upper and lower limits of water head, water demand (load), and water flow injected from a water source for node $i$, respectively. $y_{min}^G$ and $y_{max}^G$ are the upper and lower limits of head gain imposed by the pump and $f_{kmin}$ and $f_{kmax}$ are the upper and lower limits of water flow for pipe $k$, respectively.

### B. Optimal Pump Scheduling

The OPS problem aims to minimize the cost of energy consumed by the pumps while being subject to the constraints (4) to (10) [10]-[16]. This problem will determine the schedules of pumps, and the value and direction of water flow in the pipes for the second level of the purposed method.

### C. Security Injection Region

The SIR provides information about the reliability and security limits of the WDS system to customers. Besides, the water consumers, such as power plants, can easily evaluate networks' security and cope with any extreme situation. Whenever extreme events happen, all customers should control their water demands within the SIR to guarantee security and avoid any hydraulic failures. The information about water flow directions and pump status is determined by the solutions of OPS under the normal operation of WDS. Therefore, the S matrix in (4) and water flow direction in the pipes are known parameters. So, equation (4) can be modified into

$$S^*AY + S^*AH + S^*G = M(R)(f \circ f)Sgn(f)^*, \quad (11)$$

where $S^*$ and $Sgn(f)^*$ are the optimal value of $S$ and $Sgn(f)$ after OPS, respectively.

The SIR problem aims to maximize the feasible region of nodal water demands under an abnormal situation while respecting the constraint (5) to (11). The objective function of SIR will be explained in algorithm 1 in the following section.

## III. SOLUTION METHOD

### A. Convexity of Water Security Injection Region

The evaluation of SIRs is computationally hard since the mathematical model of WDSs is nonlinear and nonconvex. This paper investigates the SIR of WDSs with simple structures, i.e. radial. The method for effectively evaluating the SIR of WDSs with meshed structures will be explored in our future research. In this section, we consider two assumptions for the convexity of the WDS model established in Section II:

- First, WDS has a tree structure. Indeed, there is not any splitting node before each merging node.

- Second, all node pressure ranges are similar.

Besides, the SIR is calculated after solving the OPS, so the water flow direction in each pipe is fixed, and pump status is known.

In this section, we prove the convexity of the injection region of WDS by considering the above assumption. If any point on the line between two feasible water injection points was feasible, this set would be convex. Assume $F_a$ and $F_b$ are vectors of water flow in feasible injection profile $a$ and feasible injection profile $b$, respectively. Besides, $f_a$, $y_a$ and $h_a$ are related to the profile $a$ and $f_b$, $y_b$ and $h_b$ are related to the profile $b$. Define a $\mu \in [0,1]$ that we have $F_c = (1-\mu)F_a + \mu F_b$. In this situation, there are two lemmas:

- the $f_c = (1-\mu)f_a + \mu f_b$ is the only pipe flow related to the $F_c$.

- the pipe flow $f_c$ indicates at least one node water head pressure and level.

These two lemmas confirm that any point on the line between two feasible water injection points has only one pipe flow $F_c$ which is related to one node water head $Y_c$ and level $H_c$. These lemmas are used for proving the convexity of SIR.

For proving of first lemma, we show that $f_c$ is a pipe flow solution related to $F_c$.

$$A^T f_c = A^T((1-\mu)f_a + \mu f_b) \\ = (1-\mu)A^T f_a + \mu A^T f_b, \quad (12)$$

the values of $A^T f_a$ and $A^T f_b$ are equal to $F_a$ and $F_b$, respectively.

$$A^T f_c = (1-\mu)F_a + \mu F_b = F_c. \quad (13)$$

Then, for proving the uniqueness of this solution, assume $f_d$ is another solution related to the $F_c$. Thus, $A^T(f_c - f_d) = 0$. The first assumption implies that $A$ is a full row rank matrix so that $A^T$ is a full column rank matrix. Thus $f_d$ should be equal to $f_c$. For proving the second lemma, we use the first assumption that says the $A$ is a full row rank matrix. Thus, (4) has at least one solution. So that, $f = f_c$ admits at least one solution $Y_c$ and $H_c$.

We can prove that the SIR of a WDS with the above assumptions is a convex set according to the two lemmas. For example, the first assumption assures an existing solution for the optimization problem and helps to construct an SIR. Please refer to [22] for details about this proof.

## B. Estimating Convex Security Region

In this section, we leverage the sequential optimization algorithm proposed in [22] to construct a monotone inner polytope sequence for estimating the convex injection region of WDS based on the above assumptions. This two-step algorithm is fast since, generally, only three to four polytopes are needed to be constructed to estimate the whole region. Only the starting polytope is constructed in the first step, while the other polytopes are formed in the second step. The final convex injection region is the union of these polytopes. The proposed algorithm is detailed in Algorithm 1. This algorithm is an iterative method for approximating convex injection regions. The basic idea is that, at the $i_{th}$ inner polytope $P_i$, the furthest parallel support function for each facet is identified. The optimal points on the support functions are new vertices. These new vertices are used to expand the earlier polytope to obtain polytope $P_{i+1}$. At each iteration, this process will be repeated to achieve a more accurate feasible space. The optimization solver used in Algorithm 1 is the MATLAB nonlinear optimization with fmincon.

A two-dimensional example is given in Fig. 2 to illustrate the proposed algorithm. To achieve the first polytope, the optimization solver is applied to find the optimal values on each axis, separately. In Fig. 2, the black points show the first optimum values of the objective function, and the starting polytope is constructed with these nodes. The blue area shown in Fig. 2 (a) represents the starting polytope. In the second step of the algorithm, the optimization problem will be obtained by applying the normal direction of facet between black points, and a new constraint regarding the node water demand will be added. The optimal value of this step is shown with a blue point in Fig. 2 (b). The first polytope is constructed by connecting the blue point to black points, and the red area is added to the feasibility region. Indeed, the red area in Fig. 2 (b) is the first polytope.

**Algorithm 1** Constructing Monotone Inner Polytope Sequence

1: **Start** with inputting data such as state variables and demand variables $(d \epsilon N_d)$
2: **for** $i = 1, 2, \ldots, N_d$, generate the starting polytope, where $N_d$ is the dimension.
3: Define objective function $f_i = e_i^T d$; where $e_i$ is a unit vector.
4: Apply the optimization solver to find the optimal value of $f_i$ subject to the network constraints ((5) to (11))
5: **end for**
6: Generate the starting polytope with the optimal values
7: Let the set of $C_0$ be $C_0 = \{P_j^0 | j = 1, 2, \ldots, J\}$
8: **for** $j = 1, 2 \ldots, J$, j is the number of polytopes
9: **for** $k = 1, \ldots, N_d^{j-1}$
10: Define the new objective function $f_j^k = \vec{n}_j^k d$; where $\vec{n}_j^k$ is an outer normal direction vector of hyperplane $P_j^k$
11: Apply the optimization solver to find the optimal value of $f_i$ subject to the network constraints
12: **end for**
13: Generate the $j_{th}$ polytope by connecting the optimal values
14: Let set of $C_j$ be $C_j = \{P_j^k | j = 1, 2, \ldots, J\}$
15: **end for**

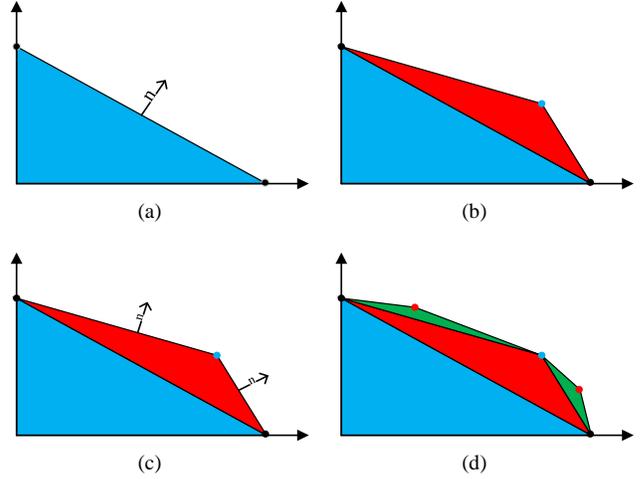

Figure 2. A two-dimensional example for the proposed algorithm

This algorithm will be repeated for constructing the next polytopes. Fig. 2 (d) shows the second polytope that is constructed by the proposed algorithm. In this Figure, the new optimal values are shown by the red points, and the green area is added to the convex security region. Repeating the iteration to infinity will result in the exact convex injection regions. However, compromising between the accuracy and the computation shows that the first few steps are generally enough for this purpose. Construction of a high dimensional sequence and analyzing its convex polytopes could be computationally complex [26]. In the real applications, the set is usually in a low dimensional space, which is particularly easy to handle by the proposed algorithm.

## IV. CASE STUDY

In this section, the proposed algorithm is implemented on two different WDSs which have different pipe lengths and diameters. In the first case study, the pump status is ON, while the pump status is OFF in the second one, as determined by the OPS solution. The main information about the test systems is provided and the simulation results are discussed.

### A. Main Information and Assumptions

The first case study is shown in Fig. 3 which is a real WDS in Iran. The water tank meets the water demand in this steady-state time because the status of the pump is OFF. The structure of the second case study, which is a 9-node WDS from the EPANET manual with some minor changes in order to meet our assumption, is shown in Fig. 4 [23]. The status of the pump in the second case study is ON, and it increases the water head pressure to overcome gravity. Some of the specifications of these systems, such as the length and diameter of pipes, are shown in Table I and Table II. The head loss coefficient of a pipe is calculated using (12) that is based on the Darcy–Weisbach equation.

$$R = f_s \frac{L}{D} \frac{1}{2gA^2} = \frac{8f_s L}{\pi^2 g D^5} \qquad (12)$$

where $L$ and $D$ are the length and diameter of the pipes, respectively, $f_s$ is the coefficient of surface resistance, and $g$ is the gravitational acceleration.

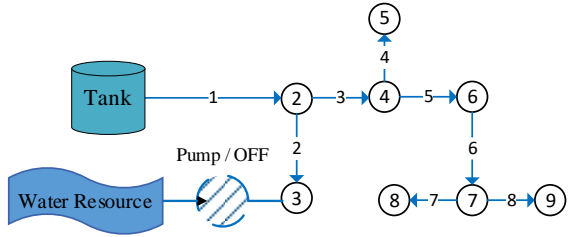

Figure 3. First water distribution network

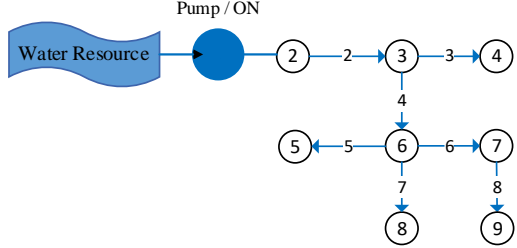

Figure 4. Second water distribution network

TABLE I. PIPE SPECIFICATION FOR TWO CASE STUDIES

| Pipe specification | | Pipe number | | | | | | | |
|---|---|---|---|---|---|---|---|---|---|
| | | *1* | *2* | *3* | *4* | *5* | *6* | *7* | *8* |
| 1 | Length (km) | 0.5 | 1 | 0.5 | 1 | 0.5 | 1 | 0.8 | 0.7 |
| | Diameter(cm) | 15 | 15 | 8 | 8 | 8 | 10 | 10 | 8 |
| 2 | Length (km) | 0.914 | 1.524 | 1.524 | 1.524 | 2.438 | 2.134 | 1.219 | 2.438 |
| | Diameter(cm) | 35.56 | 30.48 | 20.32 | 20.32 | 10.16 | 15.24 | 17.78 | 10.16 |

TABLE II. NODE SPECIFICATION FOR TWO CASE STUDIES

| Node specification | | Node number | | | | | | | | |
|---|---|---|---|---|---|---|---|---|---|---|
| | | *1* | *2* | *3* | *4* | *5* | *6* | *7* | *8* | *9* |
| 1 | Elevation(m) | 30 | 0 | 0.7 | 0.5 | 0.5 | 0.3 | 0 | 0 | 0.1 |
| | Demand (L/s) | 0 | 4 | - | 4.75 | - | 6 | 5 | 3 | - |
| 2 | Elevation(m) | 192 | 213 | 216 | 213 | 213 | 213 | 213 | 228 | 228 |
| | Demand (L/s) | 0 | 4.5 | 4.75 | 3 | - | 3.25 | 2 | - | - |

### B. Estimation of Security Injection Region

The proposed algorithm is executed on the first WDS, by considering a 3-dimensional loadability region. In the first step of the algorithm, we consider three different objective functions regarding the water demands of nodes 3, 5, and 9, and all the other water demands are fixed at the given values (mentioned in Table II). In this step, the algorithm finds the optimal values of these three nodes and constructs the starting polytope (Fig. 5 (a)). For example, the maximum acceptable value for the water demands of nodes 3, 5, and 9 are 17.25, 5.68, and 3.39, respectively. In the next step, we define a new objective function based on the proposed algorithm and apply the optimization solver to find the optimal value. The optimum value is [13.86, 0, 3.39] for the water demands of these nodes, and the algorithm constructs the first polytope based on it. Fig. 5(b) shows the first polytope for the first test system. Following the same strategy next polytopes will be constructed and added to the SIR of WDS. Fig. 5(c) and Fig. 5(d) depict the second and third polytope, respectively. The proposed strategy is implemented on nodes 5, 8, and 9 for another WDS, too. In this WDS, the water demands of nodes 5, 8, and 9 are variables, and other nodes have specific values based on Table II. The SIR of this WDS is a 3-dimensional region due to three water demand variables. Fig. 6. shows the four polytopes that are used to construct the SIR.

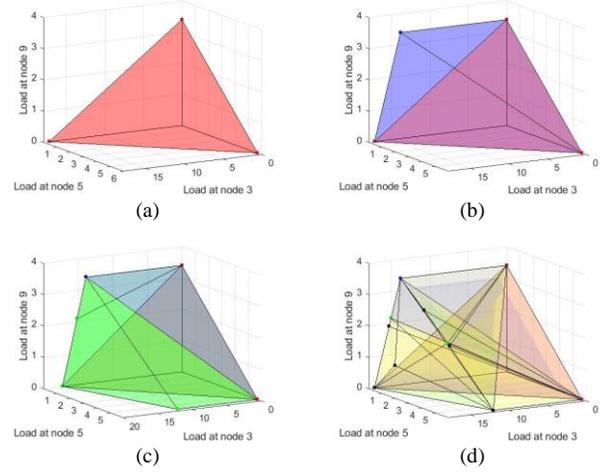

Figure 5. 3-D SIR at node 3, 5, and 9 for the first WDS. (a) the starting polytope (b) the first polytope (c) the second polytope (d) the third polytope

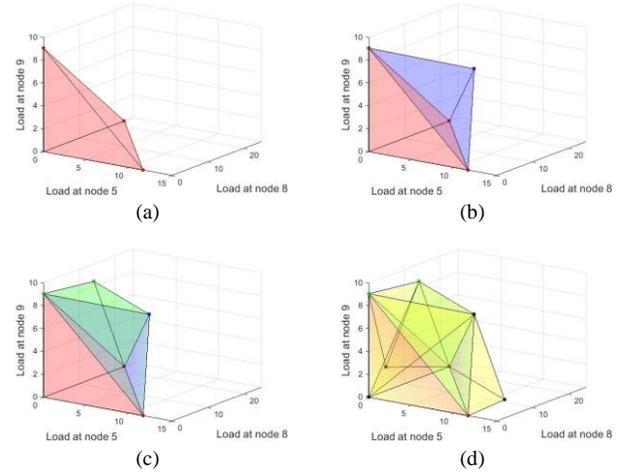

Figure 6. 3-D SIR at node 5, 8, and 9 for the second WDS. (a) the starting polytope (b) the first polytope (c) the second polytope (d) the third polytope

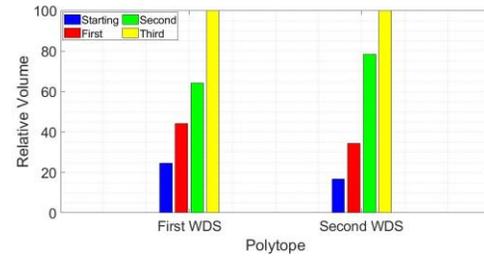

Figure 7. Relative volume of inner polytope for two WDSs

Fig. 7 depicts the relative volume of these two case studies. For example, the starting polytope relative volume for the first and second WDS is 0.2448 and 0.1667, respectively. It shows that both systems need only four polytopes to obtain the SIR.

Providing a simple and fast but accurate description of the actual SIR is the primary benefit of this algorithm. The proposed algorithm constructs the third polytope of the first WDS, that is consists of a majority of the SIR with a CPU time of 4.521 seconds (all simulations are executed in MATLAB R2019b environment with intel (R) Core (TM) i7-9700 CPU 3 GHz and 16 GB RAM.).

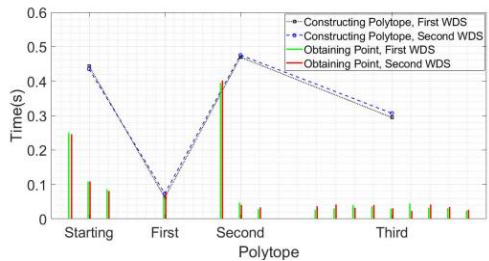

Figure 8. Excuting time of obtaining points and polytopes

Consider first WDS with three variable loads at node 3, 5, and 9 ranging in [0; 17.25], [0; 5.68], and [0; 3.39], respectively. Since each range is evenly discretized by 9 points, there are 729 3-D load combinations in total to verify. Screening these load profiles are within the third inner polytope (shown in Fig. 5 (d)) takes a relatively short time. The required time for reaching the answer by the optimization solver for each point and polytope (the starting polytope to the third one) is depicted in Fig. 8. For example, we need three points to construct the second polytope. These points are obtained in less than 0.5 seconds by applying the proposed algorithm. The optimization solver reaches the optimal values in a very short time for both of the case studies.

## V. Conclusion

This paper introduced an algorithm based on a monotone inner polytope sequence for rapidly and accurately determining the security injection region of radial water distribution systems. First, the WDS modeling is described based on the pipe network model and mass flow conservation law. Then, certain assumptions are considered to show the convexity of the injection region. When this approach is applied to case studies, the almost-whole convex region is determined only requiring three or four iterations. Furthermore, numerical studies showed that the optimization solver needs a very short time for reaching the optimal value and constructing the convex region. It proves that the proposed algorithm is an applicable and fast approach method for solving the problem.

Extending the proposed method for WDSs with more complex structures is one of the future research directions. Besides, this method can be applied to other systems like cyclic gas networks.